\documentstyle[12pt]{article}
\input epsf
\advance\leftskip by -20pt \advance\rightskip by -25pt
\advance\vsize by 4.0 truecm
\advance\voffset by -1.0 truecm
\begin{document}
\baselineskip14pt
\begin{titlepage}
\thispagestyle{empty}
\date{}
\title{Induced gravity inflation in the standard model of particle
physics}
\author{J.L.Cervantes-Cota\thanks{Electronic address:
jorge@spock.physik.uni-konstanz.de}
and
H. Dehnen\thanks{Electronic address:
dehnen@spock.physik.uni-konstanz.de} \\
University of Konstanz, p.o.box: 5560, M 678
\,\ D-78434 Konstanz }
\maketitle
\vskip 1 cm
PACS number: 98.80.Cq  \hfill astro-ph/9505069
\vskip 1cm
To appear in Nuclear Physics B.

\setcounter{page}{1}

\begin{abstract}
\baselineskip12pt

We are considering the cosmological consequences of
an induced gravity theory coupled to the minimal standard model of
particle physics.  The non-minimal coupling parameter between gravity
and the Higgs field must then be very large, yielding some new
cosmological consequences for the early Universe and new constraints
on the
Higgs mass.  As an outcome, new inflation is only
possible for very special initial conditions
producing first a short contraction era after which an inflationary
expansion
automatically follows; a chaotic inflationary scenario is
successfully achieved.   The contrast
of density perturbations required to explain the seed of astronomic
structures  are  obtained for very large values of the Higgs mass
($M_{H} >> G_{F}^{-1/2}$), otherwise
the perturbations have a small amplitude; in any case, the spectral
index of scalar perturbations agrees with the observed one.
\end{abstract}
\end{titlepage}
\setcounter{page}{2}
\section{Introduction}
\label{s1}
It is well known that inflation is a necessary constituent of modern
cosmology
in order to solve the long standing problems of standard cosmology
\cite{Gu81} as the homogeneity and isotropy
of the Universe, the horizon and flatness
problems, for reviews see \cite{KoTu90,Li90,Ol90,GoPi92}. Indeed,
inflation is desired to be accomplished by particle physics theories,
which
should be able to fulfill the cosmological standard tests
\cite{StTu84}, before
one can speak about a successfully cosmology.  However, there are
some
unwanted problems inherent to many of the inflation models like the
achievement
of the right contrast of density perturbations within a non
fine-tuning
particle physics model or enough reheat temperature
after inflation to yield the
particle content of the Universe, among others; in this sense, one
can also
speak of some ``quasi-long'' standing problems of inflationary
models.

In order to understand these issues both of cosmology and particle
physics, in
recent years it has been carried out some progress in the
experiments, like
the observations made by the Cosmic Background Explorer (COBE)
satellite
\cite{Sm92,Go94}, the possible discovery of the
top quark \cite{Ab94}, among others, as well as some theoretical
extensions of
the standard model, grand
unification theories (GUT)'s and gravity theories, see
\cite{KoTu90,Li90}.
Particularly, the cosmological consequences of induced gravity models
are
well known
\cite{Sp84,AcZoTu85,Po85,LuMaPo86,Po86,AcTr89,FuMa89,FaUn90},
but the particle physics content is still unclear, simply because the
Lagrangians used there imply scalar field associated particles
with masses greater than the Planck mass ($M_{Pl}$).  Motivated by
these
requests we have recently studied \cite{CeDe95} an
induced gravity model of inflation based on a non-minimal coupling
between
gravity and the SU(5) GUT Higgs field, as an extension to both the
minimal
SU(5) GUT and Einstein's General Relativity (GR). Now we study, what
is the
role played by the induced gravity model with a
non-minimal coupling in the standard model of particle physics,
motivated by
some aspects listed below.

The issue of inflation at the
electroweak energy scale has been recently discussed \cite{KnTu93}
motived
by the possibility that baryogenesis could take place at that scale
and also
because any net baryon number generated before would be brought to
zero by the
efficient anomalous electroweak processes, unless  the original GUT
model
was B-L conserving \cite{Gl94}.  Moreover, inflation in this energy
scale
brings other advantages: one  does not have to deal with heavy relics
or monopoles, which usually appear at GUT scales and reheating is
more
efficient.  One can also expect that the some of physics of this
scenario
could de tested in particle accelerators in the followings years.

Furthermore, recently it has been proposed induced gravity in the
standard
model could solve some other problems of particle physics and
cosmology, like
the necessity of the Higgs mass to be order of the theory cut off
\cite{Bij94} and to relate directly Mach ideas with a particle
production
mechanism by means of the equivalence principle, as well as, the
missing mass
problem \cite{DeFr93}.

In this paper, we want to discuss some early Universe consequences
of a theory of gravity coupled to the isovectorial Higgs field of the
standard
model, which is proved to produce a type of Yukawa gravitational
interaction
\cite{DeFrGh90,DeFr91,Fr92}, and which breaks down to give
rise to both some boson and fermion masses and the Newton's
constant.   Since the symmetry-breaking process of the
SU(3)$\times$SU(2)$\times$U(1) standard model is expected to
occur in the physical Universe,  we are considering inflation there.

This paper is organized as follows: in section \ref{s2} we introduce
the
theory of induced gravity in the standard model.   The cosmological
equations
for an isotropic Universe as well as the slow rollover conditions for
inflation
are presented in section \ref{s3}.  After that, in section \ref{s4}
we
analyze the two possible scenarios, i.e.,  a
modified version of new inflation as well as a chaotic inflation
model,
depending essentially on
the initial conditions for the Higgs field at the beginning  of time.
Finally,
in section \ref{s5} we stress our conclusions.

\section{Induced gravity in the standard model}
\label{s2}

We consider here an induced gravity theory coupled to the
minimal standard model of the internal gauge group
SU(3)$\times$SU(2)$\times$U(1) with the
SU(2)$\times$U(1) Higgs field $\phi$.  The Lagrange density
\cite{DeFr93}
with units $\hbar = c = k_{B} = 1$ and the signature (+,-,-,-) is:
\begin{equation}
\label{eql}
{\cal L}= \left[ \frac{\alpha}{16 \pi} \mbox{$ \phi^{\dag} \phi $}
\,\ R  + \frac{1}{2}
 \phi^{\dag}_{|| \mu} \phi^{|| \mu}
- V(\mbox{$ \phi^{\dag} \phi $}) +  L_{M} \right] \sqrt{- g}
\end{equation}
where $R$ is the Ricci scalar, the symbol $|| \mu$ means in the
following the
 covariant derivative with respect to all gauged groups
and represents in (1) the  covariant gauge derivative: $\phi_{||} =
\phi_ {| \mu } + i {\rm g} [A_ \mu , \phi]$ where
$A_{\mu} = A_ \mu {}^a \tau _a $ are
the gauge fields of the inner symmetry group, $\tau_a$  are its
generators
and ${\rm g}$ is the coupling constant of the gauge group
($|\mu $ means usual partial derivative); $\alpha $ is a
dimensionless parameter to regulate the strenght of  gravitation
and $V$ is the Higgs potential;  $ L_{M}$ contains the fermionic and
massless bosonic fields of the standard model ($L, R$ mean summation
of left-,
right-handed terms):
\begin{equation}
\label{eqlm}
L_{M}= \frac{i}{2} \bar{\psi} \gamma^{\mu}_{L,R} \psi_{|| \mu} + h.c.
- \frac{1}{16 \pi} F^{a}_{~\mu\nu} F^{~\mu\nu}_{a} -
k \bar{\psi}_{R} \phi^{\dag}  \hat{x}\psi_{L} + h.c.
\end{equation}
where $\psi$ summarizes the leptonic and hadronic Dirac wave
functions,
$F_{\mu\nu a}$ are the gauge-field strengths, $\hat{x}$ represents
the
Yukawa coupling matrix for the fermionic masses and $k$ its
(dimensionless
real) coupling constant.

Naturally from the first term of Eq. (\ref{eql}) it follows
that $\alpha~\mbox{$ \phi^{\dag} \phi $}$ plays the role of a
variable reciprocal gravitational ``constant''.   The aim of
 our theory is to obtain GR as a final effect of a symmetry
 breaking process and in that way to have  Newton's
gra\-vi\-ta\-tional
 constant $G$ induced by the Higgs field; similar theories have
 been  considered to explain  Newton's gravitational constant
in the context of a spontaneous symmetry-breaking process to unify
gravity with other fields involved in
matter interactions, but in much higher energy scales,
see Refs. \cite{Mi77,Ze79,Ad82,CeDe95}.

In this paper we want to stress the cosmological
consequences when the symmetry-breaking of the SU(2)$\times$U(1)
Higgs
field is responsible for the generation of gravitational constant as
well
as the electroweak standard particle content, i.e.,
all the fermionic masses as well as the $W^{\pm}$ -, $Z$-boson
masses.

The  Higgs potential takes the form,
\begin{equation}
\label{eqph}
V(\phi) = \frac{\lambda}{24}
\left( \phi^{2} + 6 \frac{\mu^2}{\lambda} \right)^2
\end{equation}
where we added a constant term to prevent a negative cosmological
constant after the breaking.  The Higgs ground state, $v$, is given
by
\begin{equation}
 v^2   \,\ = \,\ - \frac{6 \mu^2}{\lambda}
\end{equation}
with $V(v) = 0~$, where $\lambda$ is a dimensionless
real constant, whereas  $\mu^2~( < 0 )$ is
so far the only dimensional real constant of the Lagrangian.

In such a theory, the potential $V(\phi)$ will play the role of
a cosmological ``function'' (instead of a constant) during the
 period in which $\phi$ goes from its initial value
$\phi_o$ to its ground state, $v$, where furthermore
\begin{equation}
\label{eqg}
G \,\ = \,\  \frac{1}{\alpha v^2}
\end{equation}
is the gravitational constant to realize from (\ref{eql}) the theory
of GR
\cite{DeFr93}.  In this way, Newton's gravitational
 constant
is related in a natural form to the mass of the gauge bosons, which
for the case of the standard model one has
\begin{equation}
\label{eqmx}
M_W \,\ =  \,\ \sqrt{\pi} {\rm g} v ~~~~~.
\end{equation}
As a consequence of (\ref{eqg}) and (\ref{eqmx}) one has that the
strength parameter for gravity, $\alpha$ , is determinated by
\begin{equation}
\alpha \,\ = \,\  2 \pi \left( {\rm g}
\frac{M_{Pl}}{ M_W} \right)^2 \,\  \approx \,\ 10^{33}
\end{equation}
where $M_{Pl}=1/\sqrt{2G}$ is the Planck mass and ${\rm g} \approx
0.18$.
In this way,  the coupling between
the Higgs field and gravitation is very strong:
the fact that $\alpha>>1$ is the price paid in recovering  Newton's
gravitational constant at that energy scale and it brings some
important
differences when compared to the standard induced gravity models
\cite{CeDe95},
where to achieve successful inflation typically $\alpha << 1$
\cite{AcZoTu85,LuMaPo86}, and in that way, the existence
of a very massive particle ($> M_{Pl}$) is necessary, which after
inflation
should decay into gravitons making difficult later an acceptable
nucleosynthesis scenario \cite{BaSe8990}.

\bigskip
{}From (\ref{eql}) one calculates immediately the gravity
equations of the theory
\begin{eqnarray}
\label{eqrp}
\lefteqn{ R_{\mu \nu} -\frac{1}{2}R g_{\mu \nu} +
 {8 \pi V(\mbox{$ \phi^{\dag} \phi $})\over
\mbox{$\alpha~\phi^{\dag} \phi$}}  g_{\mu \nu} \,\ = \,\ }
\nonumber\\ &&- {8 \pi  \over \mbox{$\alpha~\phi^{\dag} \phi$}}
T_{\mu \nu}
 - {8 \pi \over \mbox{$\alpha~\phi^{\dag} \phi$}} \left[
 \phi^{\dag}_{( || \mu} \phi_{|| \nu )}   -
 \frac{1}{2} \phi^{\dag}_{|| \lambda} \phi^{|| \lambda} \,\ g_{\mu
\nu} \right]
 \nonumber\\ &&
- \frac{1}{\mbox{$ \phi^{\dag} \phi $}}
\left[ (\phi^{\dag} \phi)_{| \mu || \nu} -
(\phi^{\dag} \phi)^{| \lambda}_{~~|| \lambda} \,\ g_{\mu \nu} \right]
\end{eqnarray}
where $ T_{\mu \nu} $ is the energy-momentum tensor belonging to Eq.
(\ref{eqlm}), and the Higgs field equations are
\begin{equation}
\label{eqh}
 \phi^{|| \lambda }{}_{|| \lambda} + \frac{\delta V}{\delta
\phi^{\dagger} } -
\frac{\alpha}{8 \pi} R \phi \,\ = \,\
2 \frac{\delta L_{M} } {\delta \phi^{\dag}}  \,\ = \,\
-2 k \bar{\psi}_{R} \hat{x} \psi_{L} ~~~.
\end{equation}

In the unitary gauge the Higgs field $\phi$ takes the form, avoiding
Golstone bosons,
\begin{equation}
\label{eqsb}
\phi \,\ = \,\ v \sqrt{\mbox{$ 1+2 \chi $}} N \,\, , \,\
\mbox{$ \phi^{\dag} \phi $} = v^2  (\mbox{$ 1+2 \chi $}) N^{\dag} N =
v^{2} (\mbox{$ 1+2 \chi $}) \,\ , \quad N = const.
\end{equation}
where the new real scalar variable $\chi$ describes the
excited Higgs field around its ground state; for instance
$\phi=0$ implies $\chi = -1/2$ and
$\phi=v N$ implies $\chi = 0$.  With this new Higgs variable
Eqs. (\ref{eqrp}) and (\ref{eqh}) are now:
\begin{eqnarray}
\label{eqrx}
\lefteqn{ R_{\mu \nu} -\frac{1}{2}R g_{\mu \nu} +
\left[ {8 \pi  \over \alpha v^2} {V(\chi) \over (\mbox{$ 1+2 \chi $})
}
\right] g_{\mu \nu}
\,\ = \,\ }\nonumber\\ && - {8 \pi  \over \alpha v^2 }
\frac{1}{( \mbox{$ 1+2 \chi $} )} \, \hat{T}_{\mu \nu}
 - {8 \pi \over \alpha}  \frac{1}{(\mbox{$ 1+2 \chi $})^2}\left[
\chi_{ |\mu} \chi_{| \nu}  - \frac{1}{2} \chi_{ |\lambda}
\chi^{|\lambda}
 g_{\mu \nu} \right] \nonumber\\ &&
- \frac{2}{\mbox{$ 1+2 \chi $}} \left[ \chi_{| \mu || \nu} -
\chi^{| \lambda}_{~|| \lambda}
g_{\mu \nu} \right]
\end{eqnarray}
where  $\, \hat{T}_{\mu \nu }$ is  the {\it effective}
energy-momentum tensor
 given by
\begin{equation}
\label{eqtt}
\, \hat{T}_{\mu \nu} \,\ = \,\ T_{\mu \nu} +
\frac{(\mbox{$ 1+2 \chi $})}{4 \pi  }
{}~M_{ab}^{2}
\left( A^{a}_{~\mu} A^{b}_{\nu} - \frac{1}{2} g_{\mu \nu}
A^{a}_{~\lambda} A^{b \lambda}   \right) ,
\end{equation}
with $M^{2}_{ab}=4\pi g^{2}v^{2}N^{\dag} \tau_{(a}\tau_{b)} N$ the
gauge
boson mass square matrix.

The scalar field equation becomes, after an automatically
cancellation of
the gauge boson matter terms, a homogeneous Klein-Gordon equation,
\begin{eqnarray}
\label{eqhx}
\chi^{| \mu}_{~~|| \mu} + \frac{1}{\mbox{$(1+ \frac{4
\pi}{3\alpha})$}}
 \frac{4 \pi}{9 \alpha } \lambda v^2 ~
 \chi &  \equiv & 0
\end{eqnarray}
from which one can read immediately  the mass of the Higgs boson
$M_{H}$ , and therefore its Compton range $l_{H}$,
\begin{equation}
\label{eqm}
M_{H} \,\ = \,\
\sqrt{\frac{\frac{4 \pi}{9 \alpha}\lambda v^2}
{\mbox{$(1+ \frac{4 \pi}{3\alpha})$}}}
{}~~,~~ l_{H} \,\ = \,\ \frac{1}{M_{H} } ~~,
\end{equation}
whereby one has that the Higgs particle mass is a factor
$\sqrt{\frac{4 \pi}{3 \alpha}} \approx 10^{-17}$
smaller than the one derived from the standard model
without gravitation, for an anternative derivation
see Ref. \cite{Bij94}.  This is a very interesting property since
the Higgs mass determines the scale of the symmetry-breaking and,
moreover, $\sqrt{\lambda}/ \alpha$ shall be a very small quantity
that determines the magnitud of the density perturbations (see later
discussion).

It is worth noting that Eq. (\ref{eqhx}) has no source:
the positive trace $T$ contribution to the source turns out to be
equal in magnitude to the negative fermionic contribution, in such a
way
that they cancel each other exactly \cite{DeFr93}.  Then, no only the
gauge bosonic but also the fermionic masses no longer appear in this
equation as a source of the excited Higgs field, which
is just coupled to the very weak gravitational field contained
in the only space-time covariant derivative.  For this reason,
once the symmetry breaks down at the early Universe the Higgs
particle remains decoupled from the rest of the world, interacting
merely by means of gravity.  Therefore, it is virtually impossible
to generate the Higgs field $\chi$ or its associated particle in the
laboratory.  Consequently, its current experimental below mass
limit of $64 GeV$ does not necessarily apply here.   In fact, we
shall see
that in order to achieve successful cosmology, its value could
be much greater than this.

The energy-momentum conservation law reads
\begin{equation}
\label{eqct}
\hat T^{~\nu}_{\mu~~ || \nu} =
\frac{\chi_{| \mu}}{\mbox{$ 1+2 \chi $}} \left[ \sqrt{\mbox{$ 1+2
\chi $}}
\,\ \bar{\psi}\hat{m}\psi -
\frac{\mbox{$ 1+2 \chi $}}{4 \pi} M^{2}_{ab} A^{a}_{~ \lambda}
A^{b \lambda} \right]
= \frac{\chi_{| \mu}}{\mbox{$ 1+2 \chi $}}   \hat{T} ,
\end{equation}
where $\hat{m}=1/2 \,\ k v (N^{\dag} \hat{x} +  \hat{x}^{\dag} N )$
is the fermionic mass matrix, see Ref. \cite{DeFr93}.  The source of
this
equation is partially conterbalenced by the fermionic and bosonic
matter
fields, whose masses are acquired at the symmetry-breaking.

The potential term, which shall play the role of a positive
 cosmological
function (see the square brackets on the left hand hand of Eq.
 (\ref{eqrx})), takes in terms of $\chi$ the simple form,
\begin{equation}
\label{eqpot}
V(\chi)  \,\ = \,\  {\lambda v^4 \over 6} \chi^2 \,\ = \,\
\mbox{$(1+ \frac{4 \pi}{3\alpha})$}  \frac{3  }{8 \pi G } M_{H}^{2}
\chi^2
\end{equation}
which at the ground state vanish, $ V(\chi =0 )= 0 $.   Note
that $V(\chi) \sim  M_{Pl}^{2}  M_{H}^{2}\chi^2$ ; this
fact is due to the relationship (\ref{eqg}) to obtain GR once
the symmetry-breaking takes place.  Then, from Eq.
(\ref{eqrx}) and (\ref{eqg}) one recovers GR for the ground state
\begin{equation}
 R_{\mu \nu} -\frac{1}{2}R g_{\mu \nu} \,\ = \,\ - 8 \pi G ~ \,
\hat{T}_{\mu \nu}
\end{equation}
with the effective energy-momentum tensor, $\hat{T}_{\mu \nu}$, given
by
Eq. (\ref{eqtt}).  Newton's gravitational ''function'' is
$G(\chi)= \frac{1}{\alpha v^2}\frac{1}{\mbox{$ 1+2 \chi $}}$
and Newton's gravitational constant  $G(\chi=0)= G $.
\bigskip

Next, we proceed to investigate the cosmological consequences of such
a theory.

\section{FRW- MODELS}
\label{s3}

Let us consider a Friedman-Robertson-Walker (FRW) metric.  One has
with the use of (\ref{eqg}) that Eqs. (\ref{eqrx}) are
reduced to

\begin{eqnarray}
\label{eqap}
 \frac{\dot{a}^2 + \epsilon }{a^2} \,\ = \,\ \frac{1}{\mbox{$ 1+2
\chi $}}
\left(
{8 \pi G  \over 3 }  [ \rho  + V(\chi)  ]
- 2 \frac{\dot{a}}{a} \dot{\chi} +  {4 \pi  \over 3 \alpha}
\frac{\dot{\chi}^2}{\mbox{$ 1+2 \chi $}} \right)
\end{eqnarray}
and
\begin{eqnarray}
\label{eqapp2}
 \frac{\ddot{a}}{a}  =
\frac{1}{\mbox{$ 1+2 \chi $}} \left( {4 \pi G  \over 3 }
 [ -\rho  -3 p +2  V(\chi) ]
 -  \ddot{\chi} - \frac{\dot{a}}{a} \dot{\chi}  - \frac{8 \pi}{3
\alpha}
\frac{\dot{\chi}^2}{\mbox{$ 1+2 \chi $}} \right)  \,\ ,
\end{eqnarray}
where $a=a(t)$ is the scale factor, $\epsilon $ the curvature
constant
( $\epsilon = 0,~+1 ~or~ -1$ for a flat, closed or open space,
correspondingly),
  $\rho$ and $p$ are the matter density and pressure assuming that
the
effective  energy momentum tensor (\ref{eqtt}) has in the classical
limit the structure of that of a perfect fluid.  An overdot stands
for a time
 derivative.

In the same way Eq. (\ref{eqhx}) results in:
\begin{eqnarray}
\label{eqxp}
\ddot{\chi} + 3 \frac{\dot{a}}{a} \dot{\chi} + M_{H}^{2} \chi \,\ =
\,\ 0  \,\
{}.
\end{eqnarray}
The Higgs mass demarcates the time epoch for the rolling
over of the potential, and therefore for inflation.

The continuity Eq. (\ref{eqct}) is
\begin{equation}
\label{eqc}
\dot{\rho} + 3 \frac{\dot{a}}{a}  (\rho + p)  \,\ = \,\
{\dot{\chi} \over \mbox{$ 1+2 \chi $}} \,\ (\rho -3 p) .
\end{equation}
The matter density decreases as the Universe expands, but increases
by the mass
production due to the Higgs mechanism.
If one takes the equation of state of a barotropic fluid, i.e.
$p = \nu \rho$ with the dimensionless constant $\nu$,
Eq. (\ref{eqc}) can be easily integrated
\begin{equation}
\label{eqrho2}
\rho  \,\ = \,\
\frac{M}{a^{3 (1+\nu)}} (\mbox{$ 1+2 \chi $})^{\frac{1}{2}(1-3\nu)}
= \left\{ \begin{array}{l@{\quad if \quad}l}
  M (\mbox{$ 1+2 \chi $})^{2 }  & \nu=-1 ~~(anti-stiff ~ matter)\\
  \frac{M }{a^3} (\mbox{$ 1+2 \chi $})^{\frac{1}{2} }  &
 \nu=0~~~~(dust) \\
 \frac{M}{a^4} & \nu = 1/3 ~~(radiation) ,  \end{array} \right.
\end{equation}
where $M$ is the integration constant.   For the dust and
``anti-stiff'' matter models there is an extra factor, because of the
production mechanism; if the Higgs field is at the beginning of time
very near to its metastable
equilibrium state, $\chi \sim -1/2$, there is neither beginning
mass for the Universe nor size, see Eq. (\ref{eqao}).
For the radiation case Eq. (\ref{eqc}) is
sourceless; then there is no entropy production allowed by the
Higgs mechanism:  a radiation fluid in this theory acts as
a decoupling agent between matter and  Higgs field.
For $\nu =0$ the mass of the Universe
$M(\chi) \approx \rho a^{3} = M (\mbox{$ 1+2 \chi $})^{1/2} $
increases from zero to a final value $M$,  if the
 initial Higgs value is
\footnote{The subindex ``o'' stands for the initial
 values (at $t=0$)
of the corresponding variables.}
$\chi_{o} \approx -1/2$ (new ``inflation'');
on the other side for $\chi_{o} >1$ (chaotic inflation)
the mass decreases a factor $(\mbox{$ 1+2 \chi_{o} $})^{1/2} < 10 $
(as will become clear later), i.e., the today observed baryonic mass
should be given by $M$ if there were not extra matter production
after
 inflation.  One notes that the presence of the different types of
matter
densities  (relativistic or dust-like particles) is relevant for
the physical processes that take place, i.e., entropy production
processes, also when they bring  no important dynamical effects if
the
inflation potential dominates.

One notes that the Higgs potential is indeed
a positive cosmological function, which
corresponds to a positive mass density and a negative pressure
(see Eqs. (\ref{eqap}) and (\ref{eqapp2})), and
represents an ideal ingredient to
have inflation; that is, $V(\chi)$ shall be the ``inflaton''
potential.
But, on the other hand, there is a negative contribution to the
acceleration Eq. (\ref{eqapp2})
due to the Higgs-kinematic terms, i.e., terms involving $\dot{\chi}$
and
$\ddot{\chi}$ ;  terms involving the factor $1/ \alpha \sim 10^{-33}$
are simply too small compared to the others and can be neglected.

For inflation it is usually taken that $\ddot{\chi} \approx 0$, but
in fact
the dynamics should show up this behavior or at least certain
consistency.  For instance, in GR with the {\it ad hoc}
inclusion of a scalar field $\phi$ as a source for the inflation, one
has that at the ``slow rollover'' epoch $\ddot{\phi} \approx 0$  and
therefore $\dot{\phi} = - V^{\prime}/3H$, which implies that

\begin{equation}
\label{eqrollp}
\frac{\ddot{\phi}}{3H \dot{\phi}} =
- \frac{V^{\prime\prime}}{9 H^2} + \frac{1}{48 \pi G}
\left(\frac{V^{\prime}}{V}\right)^2 << 1 ~~~,
\end{equation}
where $H = \dot{a} /a$ is the Hubble expansion
rate (a prime denotes the derivative
with respect to the corresponding scalar field, see
Ref. \cite{StTu84}).  In the present theory, if one considers
the Higgs potential term in Eq. (\ref{eqap}) as the dominant one
\footnote{{}From now on, we shall always consider the
dynamics to be dominated
by the Higgs terms, in eqs. (\ref{eqap})-(\ref{eqxp}),
 instead of the matter density term, from which it is not possible to
drive
 inflation.} and Eq. (\ref{eqxp}) without source,
 i.e., $p = \frac{1}{3} \rho$, one has indeed an extra
 term due to the variation of Newton's ``function'' $G(\chi)$, that
is

\begin{equation}
\label{eqrollx}
\frac{\ddot{\chi}}{3H \dot{\chi}} =
\left( \frac{1}{1+\frac{4 \pi}{3 \alpha}}\right)\frac{4 \pi G}{3 }
 \left[- \frac{V^{\prime\prime}}{9 H^2} + \frac{1}{48 \pi G}
 \left(\frac{V^{\prime}}{V}\right)^2 (\mbox{$ 1+2 \chi $}) -
\frac{1}{24 \pi G}
 \frac{V^{\prime}}{V} \right] ~~~ .
\end{equation}
If $\chi<0$ the last term does not approach to zero during the
rolling
down process for $\alpha >> 1$.  Thus,
one has instead of a ``slow'', rather a ``fast'' rollover dynamics of
the
Higgs field along its potential down hill.  On the other side, for
$\chi >> 1$ there is indeed a ``slow'' rollover dynamics.

With this in mind one has to look carefully at the contribution of
$\ddot{\chi}$ : if one brings $\ddot{\chi}$ from
(\ref{eqxp}) into (\ref{eqapp2}) one
has that $M_{H}^{2} \chi$  competes with the potential
term $M_{H}^{2} \chi^2$, and during the rolling down of the
potential, when
$\chi$ goes from  $-1/2$ to $0$, $M_{H}^{2} \chi < 0$  dominates the
dynamics, and therefore instead of inflation one ends with
deflation or at least with a contraction era for the scale factor;
how
strong is the contraction era, should be determined by the set of
initial
conditions $(\chi_{o},\dot{\chi}_{o})$ .

Resuming, if one starts the Universe evolution with an ordinary new
inflation
scenario ($\chi_{o} < 0$), it implies in
this theory a ``short'' deflation instead of a ``long''
inflation period, since the Higgs field goes relatively fast to
 its minimum.  This feature should be present in theories
of induced gravity with $\alpha > 1$ and also for the BDT with
this type of potential (see for example the
field equations in Ref. \cite{Wa92}).  Considering
 the opposite limit, $\alpha<1$,
induced gravity models \cite{AcZoTu85} have proven to be successful
for inflation,
 also if one includes other fields \cite{AcTr89}; induced gravity
theories
 with a Coleman-Weinberg potential are also shown to be treatable for
 a very small coupling constant $\lambda$ with $\chi_o<0$
 \cite{LuMaPo86,Po86}, or with $\chi_o>0$ \cite{Po85}
and $\alpha < 1$ as well as $\alpha > 1$ \cite{Sp84,FuMa89,FaUn90}.
For
extended or hyperextended inflation
models \cite{LaSt89,StAc90} this problem does not arise because of
the
presence of vacuum energy during the rollover stage of evolution,
which is
supposed to be greater than the normal scalar field contribution.

With this concern one has to prepare a convenient scenario for the
Universe to begin with.  But first we would like to mention that
there are some
important aspects to be considered in the theory of the
electroweak phase transition in order to realize a more realistic
cosmological
model of inflation.  In order are the issue of the type of the phase
transition, depending on the Higgs mass, or
the right form of potential temperature correction
terms, see Refs. \cite{DiLeHuLiLid92,CaKa93}.  In the present theory,
however,
the vacuum energy is very large, $V^{1/4}\sim \sqrt{M_{Pl} M_{H}
\chi}$,
because of the gravity non-minimal coupling, which acts like a
negative mass
term to induce the phase transition; for a similar view in the
context of the SU(5) GUT see Ref. \cite{Ab81}. Therefore, one can
expect,
for a wide spectrum of Higgs mass values, the temperature corrections
to be
smaller than the contribution given by the potential Eq.
(\ref{eqph}).  Then,
in our cosmological approach, we shall achieve  the inflationary
stage just before the phase transition takes completely place, that
is, when
$\chi > 1$.  For that reason, when $\chi \sim 0$, the temperature,
shifted away
due to inflation, shall not play an important role at the Universe
dynamics.  Temperature considerations shall later be important
for entropy and particle production during reheating.

Next, we analyze the initial conditions of our models.
\section{INITIAL CONDITIONS AND INFLATION}
\label{s4}

The initial conditions we have chosen are simply
$\dot{a}_o = \dot{\chi}_o =0 $.  Equations (\ref{eqap})--(\ref{eqxp})
must
satisfy the following relations:

The size of the initial Universe is if $\epsilon =1$
\begin{eqnarray}
\label{eqao}
a^2_o = \frac{  \mbox{$ 1+2 \chi_o $} }{
  {8 \pi G  \over 3} \rho_o +
\mbox{$(1+ \frac{4 \pi}{3\alpha})$} M_{H}^{2} \chi_o^2 } ,
\end{eqnarray}
its acceleration has the value
\begin{eqnarray}
\label{eqappo}
\frac{\ddot{a}_o}{a_o}   =  \frac{1}{\mbox{$ 1+2 \chi_o $}}
 \Bigg\{ - {4 \pi G  \over 3 }
\left[  1+ \frac{1}{\mbox{$(1+ \frac{4 \pi}{3\alpha})$}}
 + \left( 1- \frac{1}{\mbox{$(1+ \frac{4 \pi}{3\alpha})$}} \right)
 3 \nu \right] \rho_o \nonumber\\
 + \mbox{$(1+ \frac{4 \pi}{3\alpha})$} M_{H}^{2} \chi_o^2 +
M_{H}^{2} \chi_o  \Bigg\}
\end{eqnarray}
and, for the Higgs field,
\begin{eqnarray}
\label{eqbx}
\ddot{\chi}_o  \,\ = \,\ - M_{H}^{2} \chi_o
\end{eqnarray}
The initial values $\rho_o$ and $\chi_o$ as well as  $M_{H}$
are the cosmological parameters to determine the  initial
conditions of the Universe.  The value of $M_{H}$ fixs
 the time scale for which the Higgs field breaks down into its ground
 state.   In order to consider  the Higgs-terms as the dominant ones
 (see footnote 2), one must choose the initial matter density
$\rho_{o} < \frac{\chi^{2}_{o}}{4 \pi}
\mbox{$(1+ \frac{4 \pi}{3\alpha})$}  M_{Pl}^{2} M_{H}^{2}$.

The question of the choice of the initial value $\chi_o$ is open: for
example, within GR for ``new inflation'' $\chi_o < 0$
\cite{Li82,AlSt82,Sta82}, whereas for ``chaotic inflation''
\cite{Li83}
$\chi_o > 0$; from the particle physics point of view one could
expect that
$\chi_{o} \sim -1/2$ at the beginning and  that it evolves to its
broken state
$\chi \sim 0$ at the end of the phase transition.  But what we know
now
is just that its actual state is the broken
one, and how this has been realized in a cosmological context remains
still
to be an open question.   This extends
the theoretical possibilities for cosmological models allowing a
mayor window
of feasible initial conditions for the scalar field responsible for
inflation.
For instance, this is of particular interest for the chaotic
inflationary
cosmology in induced gravity models \cite{FuMa89,FaUn90,CeDe95}, for
which
the initial value of the Higgs field must be far from its ground
state value, $\phi > v$.  Therefore, we are
considering both new and chaotic initial conditions, which shall
imply
different cosmological scenarios.

{\bf Scenario (a) ($\chi_{o}<0$):} {}From Eq. (\ref{eqao}) it follows
that if the initial value of the
Higgs field is strictly $\chi_{o} = -1/2$, the Universe possesses a
singularity.
If the Higgs field sits near to its metastable equilibrium point at
the beginning ($\chi_o {}^{>}_{\sim} -1/2$),
than $\chi$ grows since $\ddot{\chi}_o > 0$,
and from Eq. (\ref{eqappo}) one gets
 that $\ddot{a}_o < 0$ , i.e., a maximum point for $a_o$;
 thus at the beginning one has a contraction instead of an expansion.
Let us call this {\it rollover contraction}.

Normally it is argued that in BDT with a constant (or slowly varying)
potential producing a finite vacuum energy
density,  the  vacuum energy is dominant and is used to both to
expand the Universe and to increase the value of the scalar field.
This
``shearing'' of the vacuum energy to both pursuits  is
the cause of a moderate power law inflation instead of an
exponential one \cite{KaKaOl90}.  In this scenario the Universe
begins
with a contraction, and therefore the same shearing mechanism,
moreover
here due to the Higgs field, drives a ``friction'' process
for the contraction, due to the varying of $G(\chi)$,
 making the deflation era always weaker.
Furthermore, one can  see from Eq. (\ref{eqappo}) that the cause
of the deacceleration is the negative value of $\chi$;
then if $\dot{a} < 0$  from Eq.  (\ref{eqxp}) it
follows
$\ddot{\chi} \sim - \frac{\displaystyle \dot{a}}{ \displaystyle a}
\dot{\chi} >0$,
which implies an ``anti-friction'' for $\chi$ that tends to reduce
the
contraction, see also Ref. \cite{CaPo94} for a similar view, however,
applied
to the context of GUT's.

One may wonder if the rollover
contraction can be stopped.   As long as $\chi$ is
negative the contraction will not end, but if $\chi$ goes to positive
values, impulsed by special initial conditions, one could eventually
have
that the dynamics dominating term, $ M_{H}^{2} (\chi^2 + \chi)$, be
positive
enough to drive an expansion.  But due to the nature of Eq.
(\ref{eqxp}), if $\chi$ grows, the term
$ M_{H}^{2} \chi $ will bring it back to negative values and
cause an oscillating behavior around zero, its equilibrium state,
with an amplitude which is damped
with time due to the redshift factor $3H \dot{\chi}$.
Therefore, one has to seek  special values of $\chi_o$ ,
which will bring $\chi$ dynamically from negative values to great
enough
positive values to end up with sufficient e-folds of inflation. One
can
understand such a peculiar solution to be reached due to the
existence
of the ``inflation attractor'', for which the model of inflation is
well
behaved, see Ref. \cite{LiPaBa94} and references therein.  This
feature makes
clear that this scenario is not generic for inflation, but
depends strongly on special initial conditions; in this sense, this
is
another type of fine-tuning, which is, indeed, a ``quasi-long''
standing
problem always present by choosing the initial value of the inflaton
field in new inflationary scenarios.  For instance, in the standard
model
with a chosen Higgs mass value one finds by numerical integration
that  obtaining the required amount of inflation implies for $\chi_o$
to be
that special value given  in table \ref{t1}, see also figures 1(a)
and 2(a),
but not a very different number than this, otherwise the deflation
era
does not stop and the Universe evolves to an Einstein Universe with a
singularity; one could consider whether GR singularities are an
inevitable
consequence of particle physics.  In doing these computations,  it
has been
also assumed, of
course, that during the deflation phase the stress energy of other
fields,
e.g., radiation or nonrelativistic matter, are smaller than the Higgs
one,
otherwise an expansion follows.
\begin{figure}[ht]
\epsfxsize=6in \epsfbox{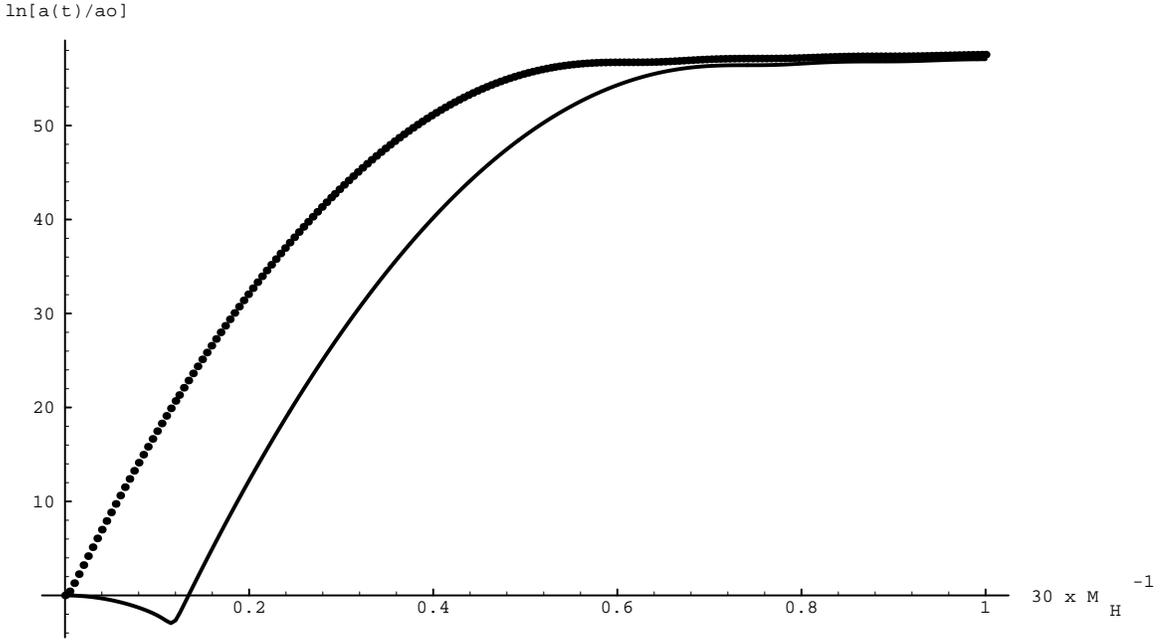}
\caption {The e-folds of inflation, $N$,
is shown for both inflationary models (a) and (b) in a
logarithmic scale.  The model (a) begins with a
``fast'' contraction followed automatically by an inflation  if
$\chi_{o} $
is that special value given in table \ref{t1}.  The upper curve
(scenario (b)) shows the behavior of inflation if
$\chi_{o} \approx 2 N _{{\rm min}}/3$ (chaotic exponential
expansion). }
\end{figure}
\vskip 1cm

{\bf Scenario (b) ($\chi_{o}>0$):}  One could  consider
initial conditions whereby
the Higgs terms $ \chi^2_o + \chi_o > 0$ dominate the dynamics to
have a minimum for $a_o$, i.e., $\ddot{a}_o > 0$, and to begin on ``a
right
way'' with expansion i.e., inflation.  That means one should start
with
a value  $\chi_o>0$ (far from its  minimum) positive enough to
render sufficient e-folds of inflation.  Thus, the ``effective''
inflaton
potential part is similar to the one proposed in the ``chaotic''
inflationary
model \cite{Li83} because of the form of the potential and the
assumed
initial Higgs value far from its potential minimum to get eventually
the
desired inflation, see also \cite{FaUn90}, but in the present theory,
of
course, we are regarding a much lower energy scale.  Then, both this
and
chaotic inflation scenarios are generic \cite{GoPi92}.

\begin{figure}[ht]
\epsfxsize=6in \epsfbox{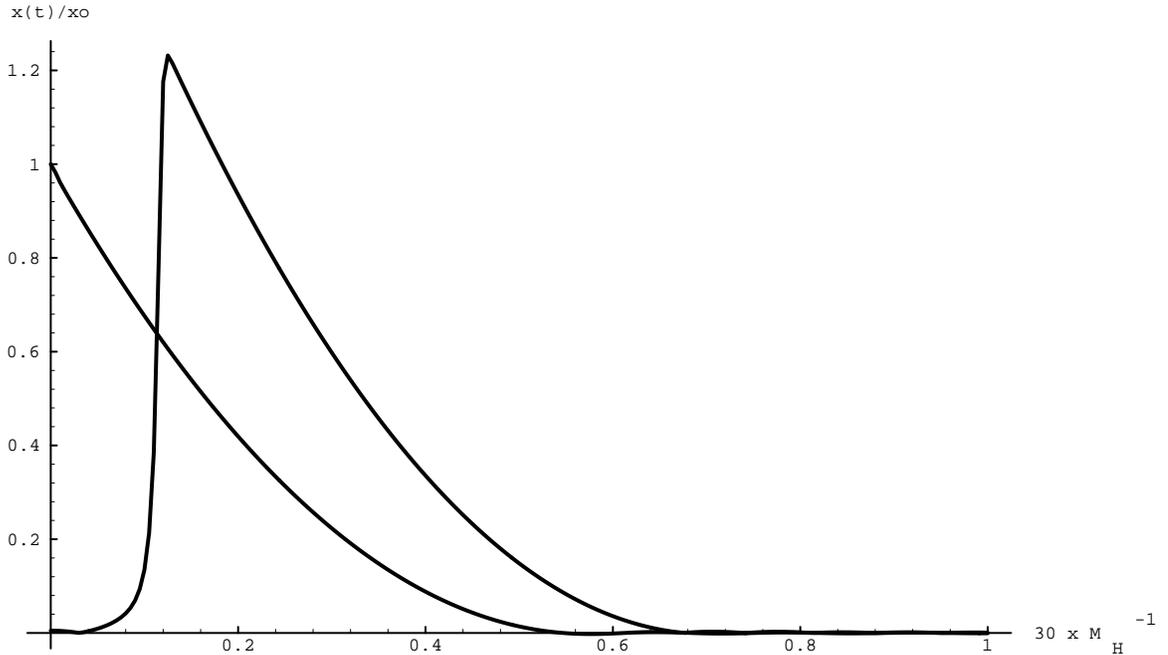}
\caption{The Higgs field of as a function of time.  (a) The Higgs
field  goes first very fast until it reaches
$\chi {}^{>}_{\sim} 2 N _{{\rm min}}/3 $;
at that point $H$ evolves faster than $\chi$, to
proceed with an inflationary phase.   (b) The same as in figure (a)
but
 here with initially
$\chi_{o} {}^{>}_{\sim} 2 N _{{\rm min}}/3 $ . The exponential
expansion
takes place directly (the ordinate axis of figure (a) is divided by
200).}
\end{figure}
\vskip 1cm

Now let us now see how the dynamics of both scenarios works: The
curvature
term $\epsilon/a^{2}$ in
(\ref{eqap}) can be neglected only after inflation began; during
the rollover contraction, scenario (a), it plays an important
role.  The terms $\frac{\dot{a}}{a} \dot{\chi}$  will be comparable
to
$8 \pi G V(\chi)/3 $
untill the high oscillation period
($H < M_{H}$) starts.  For instance, in
the chaotic scenario (b),   the slow rollover condition
$\ddot{\chi} \approx 0$ \,\ \footnote{this condition is equivalent to
both
known prerequisites
$3 H >> \ddot{\phi}/\dot{\phi},~ 3 \dot{\phi}/\phi$ of induced
gravity.}
is valid, which implies  $\dot{\chi} / \chi = - M^{2}_{H}/3H$, and
then from
\begin{equation}
\label{eqtc}
H^{2} \approx \frac{1}{\mbox{$ 1+2 \chi $}}
\left[  M^{2}_{H} \chi^{2} - 2 H \dot{\chi} \right]  ~~
\end{equation}
(with $\dot{\chi}<0$)  it follows that for $\chi > 2/3$  the Hubble
parameter
will be dominated by the potential term to have
\begin{equation}
\label{eqhh}
H \approx M_{H} \frac{\chi}{\sqrt{\mbox{$ 1+2 \chi $}}}   ~~ ,
\end{equation}
which for $\chi >>1$ goes over into $H/M_{H} \sim \sqrt{\chi /2}>>1$,
giving
cause for the slow rollover chaotic dynamics.   Indeed, the rollover
time
is $\tau_{{\rm roll}} \sim 3 H/M^{2}_{H} $ , i.e.,
\begin{equation}
\label{eqroll}
N = H \tau_{{\rm roll}} \,\ \approx \,\ 3 H^{2}/M^{2}_{H} \,\ \approx
\,\
 3 \frac{\chi^{2}}{\mbox{$ 1+2 \chi $}}  ~~ ,
\end{equation}
yielding  enough e-folds of inflation ($N$) requires to choose the
initial
Higgs field value sufficient great, and the slow rollover conditions
asures
an inflationary stage; for $\chi_{o} >>1 $ it follows that $ \chi_{o}
{}^{>}_{\sim} 2 N_{{\rm min}}/3$;
this value can be checked by numerical integration, see figure 1(b)
and 2(b).
Note that the required amount of inflation depends only on the
initial value
of $\chi_{o}$ through $ N {}^{>}_{\sim} 3 \chi_{o}/2 $, that is,
$H/M_H$ does not depend on the energy scale of inflation but on the
initial value $\chi _o$; in other words, enough inflation is
performed
automatically and independently of $\alpha$ as was pointed out in
Ref.
\cite{FaUn90}.  Moreover, it is well known that at higher energy
scales the
amount of e-folds required for successfully inflation is bigger. In
this
model one can see this as follows: suppose that at the time
$t_{*} \approx 10^{l} M^{-1}_{H}$
reheating (RH) takes place and assume that
$T_{RH}=\sqrt{M_{Pl} M_{H}}$ (we use this simple relationship, but a
properly
account of reheating is developed in Ref. \cite{KoLiSt94}) then one
has that
\begin{equation}
\label{eqnmim}
N_{{\rm min}} = \frac{1}{3} {\rm ln} S -
\frac{1}{2} {\rm ln} \frac{M_{Pl}}{M_{H}} +
\frac{1}{2} {\rm ln} \frac{2}{\chi_{o}} - \frac{2 l}{3} {\rm ln} 10
\end{equation}
where $S$ is the entropy of the Universe. Therefore the value of
$N_{{\rm
min}}$
should be also greater by increasing the energy scale of inflation.
For
instance, for $S=10^{88}$  and $M_{H}=10^{-5} M_{Pl}$ (see later
discussion)
one obtains the $N_{{\rm min}}\approx 57$.  In table \ref{t1} are
computed the
$N_{{\rm min}}$ values for three Higgs mass values.

On the other side if $\chi_{o}$  is negative,   the {\it rollover
contraction} phase in scenario (a) happens, but in this case Eq.
(\ref{eqtc})  indicates $ H/M_{H} \approx |\chi| < 1$, that is, the
scale factor evolves slower than the Higgs field; and for special
values of
$\chi_o $, the Higgs field evolves to values greater or equal than
$2 N_{{\rm min}}/3$ to gain conditions very similar to scenario (b),
see
figures 1 and 2. In table \ref{t1} are given the initial Higgs field
values when successfully inflation is achieved in scenario (a)
for different Higgs mass values. The very special initial value
$\chi_{o}$
depends also only on $N_{{\rm min}}$; again
because of the fact that at higher energy scales
$N_{{\rm min}}$ is greater, then $\chi_{o}$ is slightly
different for the various Higgs masses in table \ref{t1}.

Summarizing, for the two possibilities of Universe's models, one
has the following: In the chaotic scenario (b), the initial value
should be
$\chi_{o} > 2 N_{{\rm min}}/3$ in order to achieve sufficient e-folds
of
inflation.  And in the scenario (a), only for special initial
values of the Higgs field, the Universe
undergoes a small contraction  which goes over automatically into a
sufficiently long inflation period;  otherwise, for other initial
negatives
values of $\chi_{o}$, the Universe contracts to a singularity.   The
cosmological model integrated and shown in the figures correponds to
 $M_{H}=10^{14} GeV$ (see discussion later).  For the other Higgs
mass values,
the dynamics is very similar, giving an output resumed in table
\ref{t1}.

At the end of inflation the Higgs field begins to oscillate with a
frequency
$M_{H} > H$ and the numerical solution goes smoothly into an
oscillation dominated Universe, reaching  a normal Friedmann
regime \cite{Tu83}.  This can
be seen as follows:  First when  $H {}^{>}_{\sim} M_{H}$   with
$H \approx {\rm const.}$, $~ \chi \approx e^{-3H/2~t} cos M_{H} t$ is
valid,
later on when
$H << M_{H}$, $H\sim 1/t$ and $ \chi \sim 1/t ~ cos M_{H} t $ give
rise to
$a \sim t^{2/3}$, i.e., a matter dominated Universe with  coherent
oscillations, which will hold on if the Higgs bosons
do not decay; in figures 3 and 4  the
behavior of the scale factor and the Higgs field is shown until the
time
$100 M^{-1}_{H}$; the numerics fit very well the  ``dark''
 matter dominated solutions.  Let us consider this possibility more
in
detail: then, the average over one oscillation of the absolute value
of the
effective energy density of these oscillations,
$\rho_{\chi} {}^{>}_{\sim} V(\chi) = \frac{3}{4 \pi} M^{2}_{Pl}
M^{2}_{H}
\chi^{2}$,  is such that
\begin{equation}
\label{eqrxosc}
\frac{\rho_{\chi}}{\rho_{\chi_{\rm osc}}} =
\left( \frac{t_{{\rm osc}}}{t} \right)^2
\end{equation}
where $t_{{\rm osc}}$ is the time when the rapid oscillation
regime begins.  From Eq. (\ref{eqrxosc})  one can compute the present
(labeled with a subindex $n$) energy density of these oscillations if
they
were to exist.  Then,
\begin{equation}
\label{eqrxn}
\rho_{\chi_{n}} {}^{>}_{\sim}
\frac{3}{4 \pi} M^{2}_{Pl} M^{2}_{H}  \chi^{2}_{{\rm osc}}
\left( \frac{t_{{\rm osc}}}{t_{n}} \right)^2 .
\end{equation}
{}From the figures it is evident that $t_{{\rm osc}} = 20 M^{-1}_{H}$
and
$\chi_{{\rm osc}} = 10^{-2}$, and
$t_{n}\sim 10^{17} {\rm s}= 1.5\times10^{41} GeV^{-1}$  implying,
for all Higgs mass values chosen in table \ref{t1}, that
$\rho_{\chi_{n}} {}^{>}_{\sim} 3 \times 10^{-47} GeV^{4}$,
i.e., the Higgs oscillations could solve the missing mass
problem of cosmology, implying the existence of cold dark matter,
since after some time
as the Universe expands the Higgs particles will have a very slow
momentum
owing to their big mass.  Furthermore,  if some amount of Higgs
oscillations
decays into relativistic particles with $\rho \sim 1/a^{4}(t)$, they
can
dominate the dynamics of the Universe only for a time era, until the
density
of the remanent oscillations (if they are still there), decreasing as
$1/a^{3}(t)$, govern again the scale factor evolution, giving place
also in
this case to a dark matter dominated Universe, even if the remanents
strongly
interact with each other \cite{KoLiSt94}.

\begin{figure}[ht]
\epsfxsize=6in \epsfbox{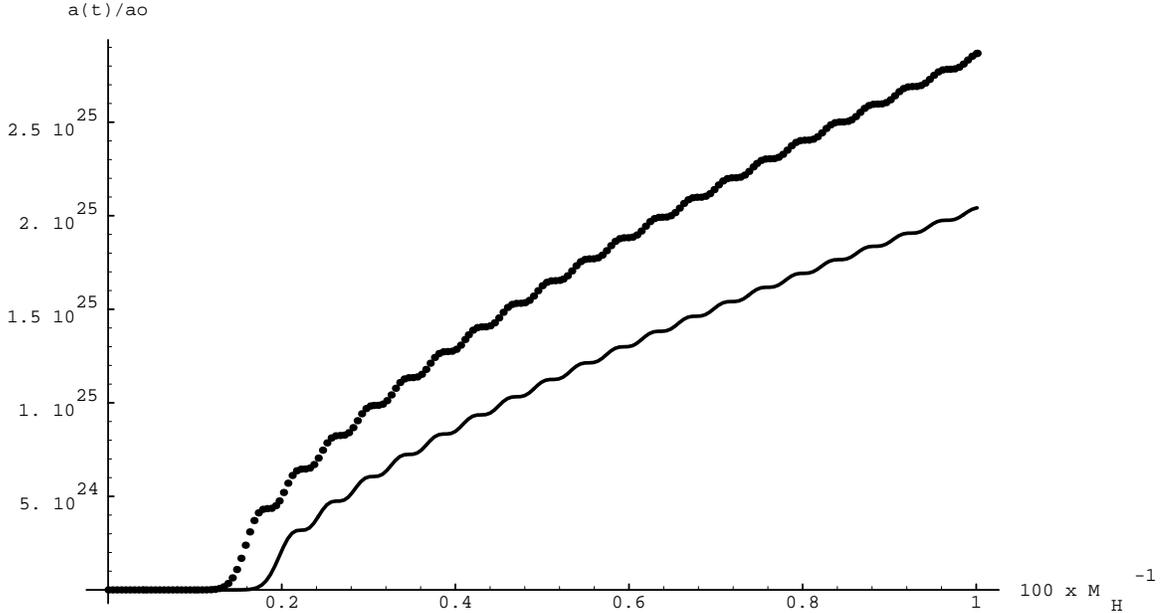}
\caption {Again the scale factor evolution as in figure 1, but now
till
$t=10^{2} M^{-1}_{H}$.  One notes that the inflation time is
approximately
$t=20 \times M^{-1}_{H} $, later on, the Universe
is ``dark'' matter dominated, perhaps until today, if
reheating didn't take away the coherent  Higgs oscillations.  It can
be
seen the track imprinted by the Higgs coherent oscillations in the
scale
factor evolution at that time scale; later on, this influence will be
imperceptible. }
\end{figure}
\vskip 1cm

\begin{figure}[ht]
\epsfxsize=6in \epsfbox{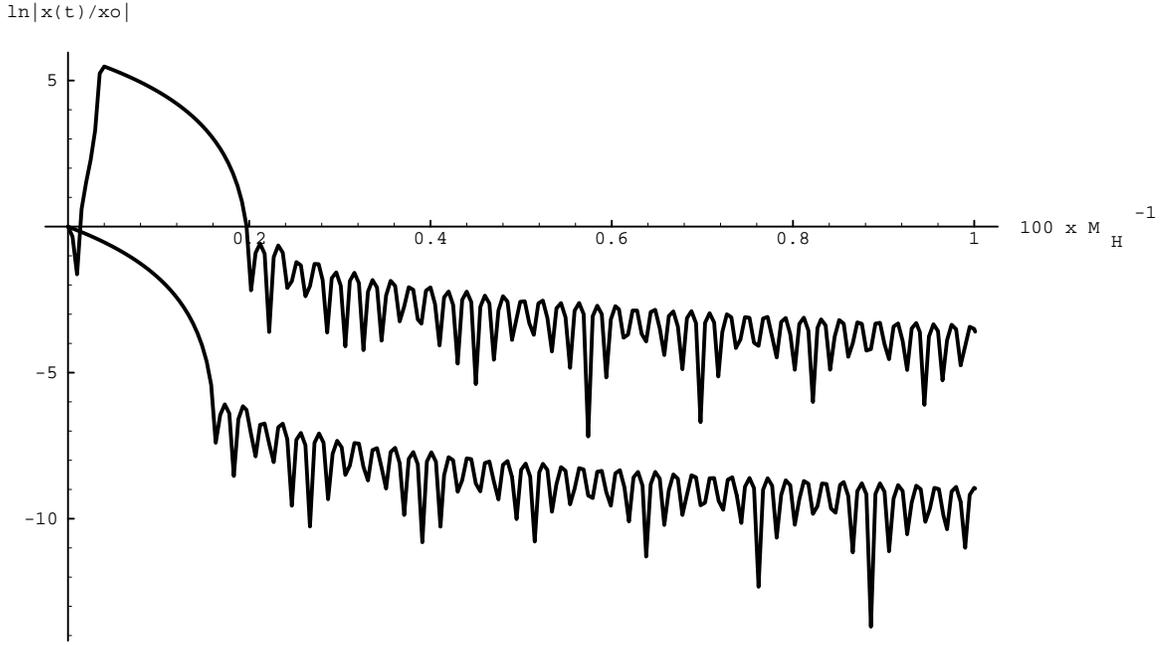}
\caption{The evolution of the Higgs oscillations is shown in
logarithmic scale during and after inflation.  In scenario (a) the
Higgs field
jumps from very small values to $2 N _{{\rm min}}/3$ to achieve
inflation,
later it begins to oscillate.  In scenario (b), the Higgs field
disminishes
until it begins to oscillate.}
\end{figure}
\vskip 1cm

There is, however, a very important problem: if one tries to explain
the today
observed baryonic mass of the Universe, given by
$M(t) \approx M (\mbox{$ 1+2 \chi $})^{\frac{1}{2} (1-3\nu)}/a^{3
\nu}$, one
has that after inflation it is too small unless the perfect fluid
behaves like
dust particles ($\nu=0$); moreover, the temperature at that
time should also be too small.  For solving these problems
one has to assume that some amount of the Higgs oscillations decay
into
baryons and leptons.
At the time around $t_{*}$ the Higgs field should decay into
other particles  with a decay width $\Gamma_H$ to give
place to a normal matter or radiation Universe expansion, producing
the reheating of the Universe \cite{AlTuWi82,DoLi82,AbFaWi82}.  If
reheating
takes place, the still remaining energy of the scalar field at
$t_{*}$ is
converted into its decay products.  This would mean that the
cosmological
``function'' disappears to give then rise to  the known matter
of the Universe.   But if  coherent oscillations still stand, they
are
the remanents of that
cosmological ``function'', which is at the present  however
 invisible to us in the form of cold dark matter, moreover, the Higgs
particle does not interact with the rest
of the particles but only gravitationally, therefore, the Higgs
oscillations
don't change baryogenesis and/or nucleosynthesis processes.
Now suppose that the oscillations really did decay.  Mathematically,
the way of
stopping the oscillations or to force the decay is to introduce a
term
$\Gamma_{H} \dot{\chi} $ in Eq. (\ref{eqxp}).  The Universe  should
then
reheat up to the temperature $T_{RH} \approx \sqrt{M_{Pl}
\Gamma_{H}}$ where
 $\Gamma_{H}$ depends, of course, on the decay products, see however
Ref.
\cite{KoLiSt94} for a more realistic reheating scenario.  For
example, if the
coherent oscillations decay into two light fermions, it is valid that
$\Gamma_{H} \approx g^{2} M_{H}$; for the reheating this would mean
that
$ T_{RH} \approx g \sqrt{ M_{H} M_{Pl} }$
which should be enough for non-GUT baryogenesis to occur \cite{Do92},
for the given Higgs mass values of table \ref{t1}.
But one should be aware of the production of gravitational
radiation  as a decay output of the oscillations \cite{BaSe8990};
however,
 it could be also possible that other decay
channels are  important, since the symmetry-breaking takes
here place at a much more smaller energy scale than the Planck one.

\vskip 0.5cm

{\renewcommand{\arraystretch}{2.5}\begin{table}\caption{The Higgs
field initial values yielding sufficient inflation for three Higgs
masses in
both new and chaotic scenarios.  We have overall taken
$\dot{\chi}_{o}=0$.}
\begin{displaymath}
\begin{array}{cccc}\hline\hline
{\rm Higgs \,\ mass} & \quad\quad  {\rm e-folds}\quad\quad
&{\rm new \,\ inflation}&{\rm chaotic \,\ inflation}\\[-4ex]
 M _{H}&N _{{\rm min}}&\chi _{o}& 2N _{{\rm min}}/3 \\ \hline
10 ^{14}GeV & 57 & -0.155088 & 38\\ \hline
10 ^{2}GeV & 44 & -0.155061 & 29\\ \hline
10 ^{-6}eV & 24 & -0.155050 & 16 \\ \hline \hline
\nonumber\end{array}
\end{displaymath}\label{t1}\end{table}}

\vskip 0.5cm

The contrast of density perturbations $\delta\rho/ \rho$ can be
considered
in scenario (b), or in scenario (a)  when $\chi$ has evolved to its
maximal
value to have very similar slow rollover conditions as in (b). Then
one has
\cite{FaUn90,MaSa91}
\begin{equation}
\label{eqc1}
\frac{\delta\rho}{\rho} \;\vline {\atop{ \atop {}_{t_{1}}}} \approx
\frac{1}{\sqrt{1+\frac{3\alpha}{4 \pi}}}
H \frac{\delta\chi}
{\mbox{$\dot{\chi}$}} \;\vline {\atop{ \atop {}_{t_{1}}}} =
\sqrt{\frac{3}{\pi \alpha}}
\left( 1+\frac{4 \pi}{3 \alpha} \right)
\frac{M_{H}}{v}\frac{\chi^{2}}{\mbox{$ 1+2 \chi$}}
\;\vline {\atop{ \atop {}_{t_{1}}}}
 \end{equation}
where $t_{1}$ is the time when the fluctuations of the scalar
field leave $H^{-1}$ during inflation.  At that time, one finds that
\begin{equation}
\label{eqc2}
\frac{\delta\rho}{\rho} \;\vline {\atop{ \atop {}_{t_{1}}}} \approx
\frac{2}{\sqrt{3}}
\frac{\sqrt{\lambda}}{\alpha}
\frac{\chi^{2}}{\mbox{$ 1+2 \chi$}} \;\vline {\atop{ \atop
{}_{t_{1}}}}
\approx
\sqrt{\frac{1}{6 \pi}}
\frac{M_{H}}{M_{Pl}} ~ N(t_{1}) \,\ \approx \,\
 10 \frac{M_{H}}{M_{Pl}} < 10^{-4} \-- 10^{-5} \,\ ,
\end{equation}
where we have used $\left( 1+\frac{4 \pi}{3 \alpha} \right) \approx
1$
and we recall that
$N(t_{1})=\frac{\displaystyle 3 \chi^{2}}{\mbox{$ 1+2 \chi$}}
\;\vline {\atop{ \atop {}_{t_{1}}}}$
from Eq. (\ref{eqroll}), which can be numerically checked from figure
4.
In order to have an acceptable value of $\delta\rho/\rho\approx
10^{-5}$
one is forced to choose $M_{H} < 10^{-5} \-- 10^{-6} M_{Pl}$.  In
that way
the magnitud of density perturbations can give rise to the
observed astronomic structures, corresponding approximately to
$N(t_{1})=50$
e-folds before inflation ends. The accomplishment of the right
density contrast
at this energy scale determines a very large value for $\lambda >>
1$,  making
a tide interaction at the outset of inflation; this huge value
for $\lambda$ brings the energy scale of inflation to be
approximately
as great as the GUT's inflationary scenarios; in the induced gravity
model
considered in Ref. \cite{CeDe95} Eq. (\ref{eqc2}) also holds, that
is,
this equation is a caracteristic of all induced gravity models,
as was also pointed out in Refs. \cite{FaUn90,MaSa91}.  If one
assumes that
$\lambda \approx \alpha$, the Higgs mass becomes of the order of
magnitud of
the electroweak scale ($G^{-1/2}_{F}$), then, this
theory is equivalent to a massive Yang-Mills theory,
which is in agreement with the present experiments
even though it is non renormalizable, because the cut-off
dependence is only logarithmic  \cite{Bij94}. On the other hand, if
$\lambda \sim 1$ implies $M_H \approx 10^{-6} eV$
and hence  inflation should be realized at approx.
$t \sim 10^{-9} {\rm s}$.  This could also be possible,
but then $\delta\rho/\rho$ is extremely small, leaving the structure
formation
problem aside from inflation.  In the latter case, the reheating
temperature is
about $10^{2} GeV$, $i.e.$, on the limit for non-GUT
baryogenesis to occur \cite{Do92}.

The spectral index of the scalar perturbations, $n_{s}$,
serves has a test for models of the very early Universe,
indenpendently of the
magnitud of the perturbations and can be calculated, using the slow
roll
approximation up to second order \cite{KoVa94,LiPaBa94}, however, for
$\alpha>>1$,
one can just take the first order to be sufficiently accurate
\cite{Ka94}:
\begin{equation}
\label{eqns}
n_{s}= 1 - \frac{2 \alpha}{N ~\alpha + \pi} \approx 1 - \frac{2}{N}
\,\ ,
\end{equation}
for $N=50$, it implies $n_{s}\approx 0.96$ in accordance with the
recent
COBE DMR results \cite{Go94}.

The perturbations on the microwave background
temperature are also well
fitted.  The gravitational wave perturbations considered normally
must also be very small \cite{AcZoTu85},
\begin{equation}
\label{eqgw}
h_{GW} \approx \frac{H}{M_{Pl}} \approx \frac{M_{H}}{M_{Pl}}
\sqrt{ \frac{\chi}{2} } \;\vline {\atop{ \atop {}_{t_{1}}}} < 10^{-5}
\,\ ,
\end{equation}
for the all above mentioned Higgs mass values.
\section{CONCLUSIONS}
\label{s5}

We have presented the induced gravity model coupled to the standard
model
of particle physics, where the cosmological inflaton is precisely the
SU(2)
isovectorial Higgs field. As a consequence of this, one has some new
features
for both particle physics and cosmology. It was shown that the
combination of the fundamental masses of the theory, due to the
non-minimal
gravity coupling, in a natural way fixes $\alpha$ to be $\sim
10^{33}$ and
the Higgs mass to be $\sqrt{\frac{4 \pi}{3 \alpha}}$ less than that
in the
standard SU(2) theory.  Indeed, the excited Higgs decouples from the
other particles and interacts just by means of the very weak
gravitational field contained in the space-time covariant derivative.
Also, because of the non-minimal coupling the vacuum
energy responsible for inflation is $V^{1/4}\sim \sqrt{M_{Pl} M_{H}
\chi}$,
bringing the energy scale of inflation equal to that of the Higgs
mass, by
means of Eq. (\ref{eqhh}).  The cosmological equations
(\ref{eqap}-\ref{eqc})
present for $\chi_{o}<0$ a  {\it rollover contraction} era in
scenario (a), whereby only for special initial values the model can
evolve to
its ``inflaton attractor'' giving rise, after all, to a chaotic
scenario, where
inflation takes place by the virtue of a normal rollover
approximation.

After inflation, the universe is oscillation dominated, and
without its total decay one could explain the missing mass problem of
cosmology given today in the form of cold dark matter.

As a matter of fact, the cosmological model cannot explain by itself
the today  observed baryon mass of the universe, for which one is
forced to
 look for a reheating scenario after inflation.  A
carefully treatment of it is not developed here, but elsewhere
\cite{KoLiSt94};
nevertheless it was point out the reheating Temperature should be
enough
for non-GUT baryogenesis to occur. However,
the question whether to much gravitational radiation is generated to
eventually spoil a normal nucleosynthesis procedure remains open at
this
 energy scale.

The right amplitude of scalar and tensor density perturbations
required to
explain the seeds of galaxy formation imposes very great values
to the Higgs mass,
$M_{H} < 10^{-5} \-- 10^{-6} M_{Pl}$, otherwise, inflation at lower
energy
scales does not account for solving that problem.  At any energy
scale,
induced gravity models predict a value of the spectral index,
$ n_{s} \sim 1 $, according with the recent observations,
see Refs. \cite{Ka94,Go94}.

As a final comment we would like to point out that the induced
gravity
model in the SU(5) theory seems to be better accomplished because,
in that case, the ratio of the Higgs to Planck mass is in a natural
way
of the order of $10^{-5}$ \cite{CeDe95}, achieving right perturbation
amplitudes.  Contrary to that case, the present SU(2) Higgs gravity
inflationary model requires unnatural big Higgs mass values in order
to render
a successfully cosmology. This is, somehow, the price paid in
matching gravity
to a very low energy scale; it reminds us once more of a quasi-long
standing problem of inflation:
whereas cosmology is happy, particle physics is infelicitous, or
inversely.

{\bf{Acknowledgment}}
One of the authors (J.L.C.) acknowledges DAAD and CONACyT (reg.
58142)
for the grant received since without it the  work would not
have been achieved.


\end{document}